\documentclass[12pt,prb]{revtex4}
\usepackage[T1]{fontenc}
\usepackage[latin9]{inputenc}
\usepackage[a4paper]{geometry}
\usepackage{color}
\usepackage{amsmath}
\usepackage{amssymb}
\usepackage{graphicx}

\makeatletter

\providecommand{\tabularnewline}{\\}

\@ifundefined{textcolor}{}
{%
 \definecolor{BLACK}{gray}{0}
 \definecolor{WHITE}{gray}{1}
 \definecolor{RED}{rgb}{1,0,0}
 \definecolor{GREEN}{rgb}{0,1,0}
 \definecolor{BLUE}{rgb}{0,0,1}
 \definecolor{CYAN}{cmyk}{1,0,0,0}
 \definecolor{MAGENTA}{cmyk}{0,1,0,0}
 \definecolor{YELLOW}{cmyk}{0,0,1,0}
}

\makeatother

\begin{document}

\title{Electron paths and double-slit interference in the scanning gate
microscopy}

\author{K. Kolasi\'{n}ski}

\address{AGH University of Science and Technology, Faculty of Physics and
Applied Computer Science,\\
 al. Mickiewicza 30, 30-059 Kraków, Poland}

\author{B. Szafran}

\address{AGH University of Science and Technology, Faculty of Physics and
Applied Computer Science,\\
 al. Mickiewicza 30, 30-059 Kraków, Poland}
\begin{abstract}
We analyze electron paths in a solid-state double-slit interferometer
based on the two-dimensional electron gas and their mapping by the
scanning gate microscopy (SGM). A device with a quantum point source
contact of a split exit and a drain contact used for electron detection
is considered. We study the SGM maps of source-drain conductance ($G$)
as functions of the probe position and find that for a narrow drain
the classical electron paths are clearly resolved but without any
trace of the double-slit interference. The latter is present in the
SGM maps of backscattering ($R$) probability only. The double-slit
interference is found in the $G$ maps for a wider drain contact but
at the expense of the loss of information on the electron trajectories.
Stability of $G$ and $R$ maps versus the geometry parameters of
the scattering device is also discussed. We discuss the interplay
of the Young interference and interference effects between various electron paths introduced by the
tip and the electron detector.
\end{abstract}
\maketitle

\section{Introduction}

The scanning gate microscopy (SGM) is an experimental technique \cite{sgmr1,sgmr2}
that probes transport properties of devices with the two-dimensional
electron gas -- buried shallow beneath the surface of the sample --
by the charged tip of the atomic force microscope. The technique has
been used in particular for investigation of quantum point contacts
-- the most elementary quantum transport devices \cite{topinka1,topinka2}.
The conductance maps gathered with the SGM contain characteristic
oscillation of the period of half the Fermi wavelength that appear
due to the interference of electron wave functions incoming from the
quantum point contact and backscattered by the tip \cite{int1,int2,int3,int4,Jalabert2010,Gorini2013,Brun2014,kozikov2013,paradiso}.
In our recent paper \cite{Kolasinski2014} we have proposed a system
with split source channel for observation of the double-slit interference.
The Young interference should be present in the SGM maps provided
that the transport in the channel which feeds the split source occurs
in the lowest subband of lateral the quantization \cite{Kolasinski2014}.
The proposal of Ref. \cite{Kolasinski2014} and most of the experimental
studies \cite{int1,int2,int3,int4} dealt with systems in which the
electron after passing through the constriction defining the quantum
contact enters an infinite half-plane. In this work we consider the
possibility of observation of electron paths in the context of the
double-slit interference. According to the which-path thought experiment
by Feynman \cite{Fey} determination of the slit the electron goes
through destroys the double-slit interference. Here, we demonstrate
that SGM can be used for detection of classical electron paths --
although without indicating the one taken by the electron -- with
a simultaneous resolution of the double slit interference effects.
The SGM was early used for detection of the semi-classical electron
trajectories as deflected by the Lorentz force \cite{crook}, including
observation of the skipping orbits \cite{skipping,aidala} for systems
with an additional confinement. Observation of electron paths requires
placing an electron detector in the system -- usually another QPC
serving as the drain channel \cite{crook,skipping,aidala}. In the
present paper we consider a system with a split QPC serving as the
electron source and the second QPC used for detection of the electron
passage. We demonstrate that the source-drain conductance maps for
a narrow drain detector indicate the classical paths but miss the
double-slit interference. For enlarged drain width the double-slit
interference appears in the map but the image of the paths is lost.
The simultaneous observation of both the semi-classical paths and
the Young interference is possible when besides the map of source-drain
conductance one considers the SGM maps for backscattered electrons
-- or equivalently -- the maps for conductance between the source
and the rest of the system excluding the drain detector. We also discuss
the electron paths in terms of the quantum consequences in various
interference effects. In particular we also point out that backscattering
by the tip induces interference between the slits with a characteristic
SGM pattern that is superimposed on the Young interference image.

\section{Model}

\begin{figure*}
\begin{centering}
\begin{tabular}{lc}
\includegraphics[width=0.3\paperwidth]{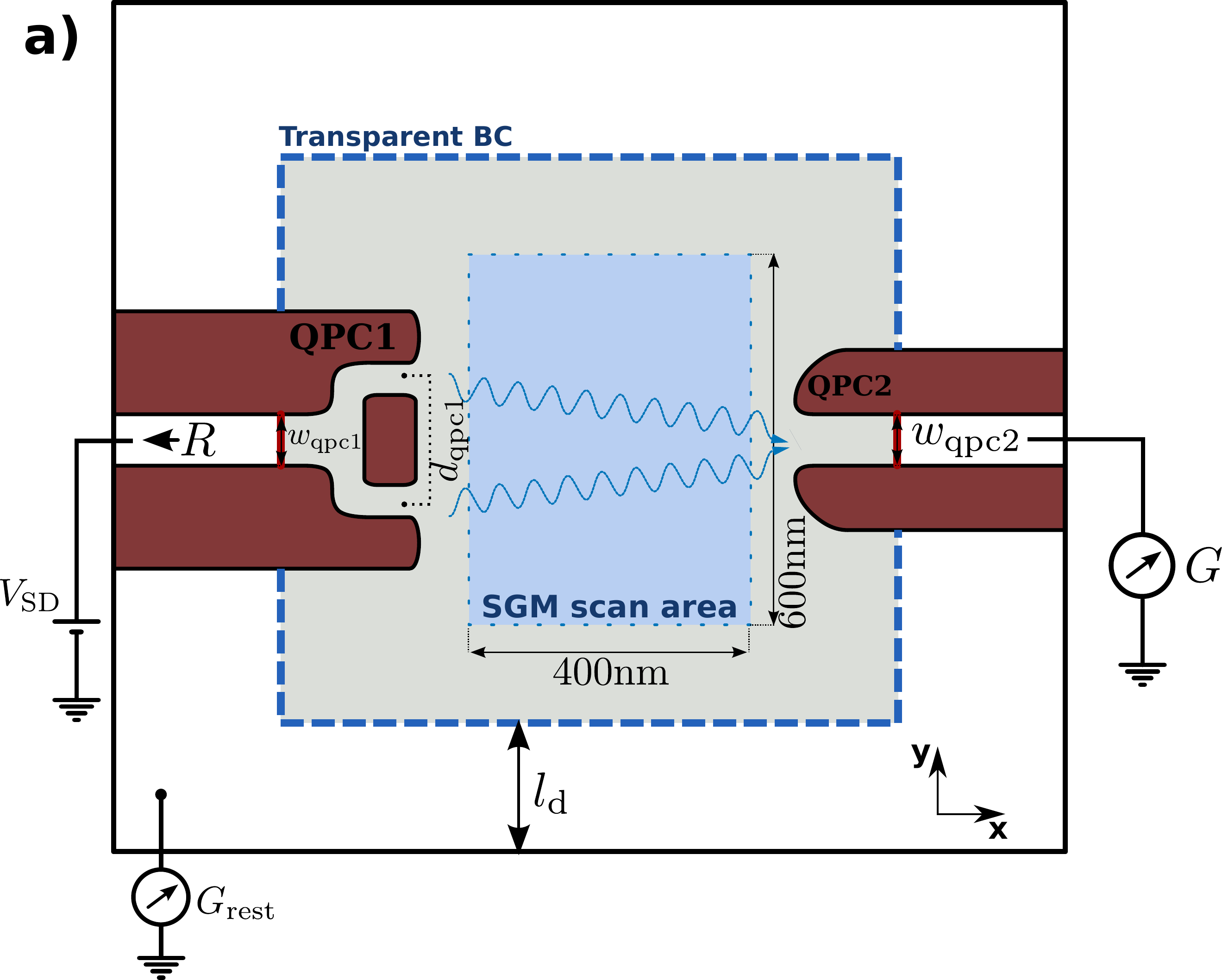}  & \includegraphics[width=0.5\paperwidth]{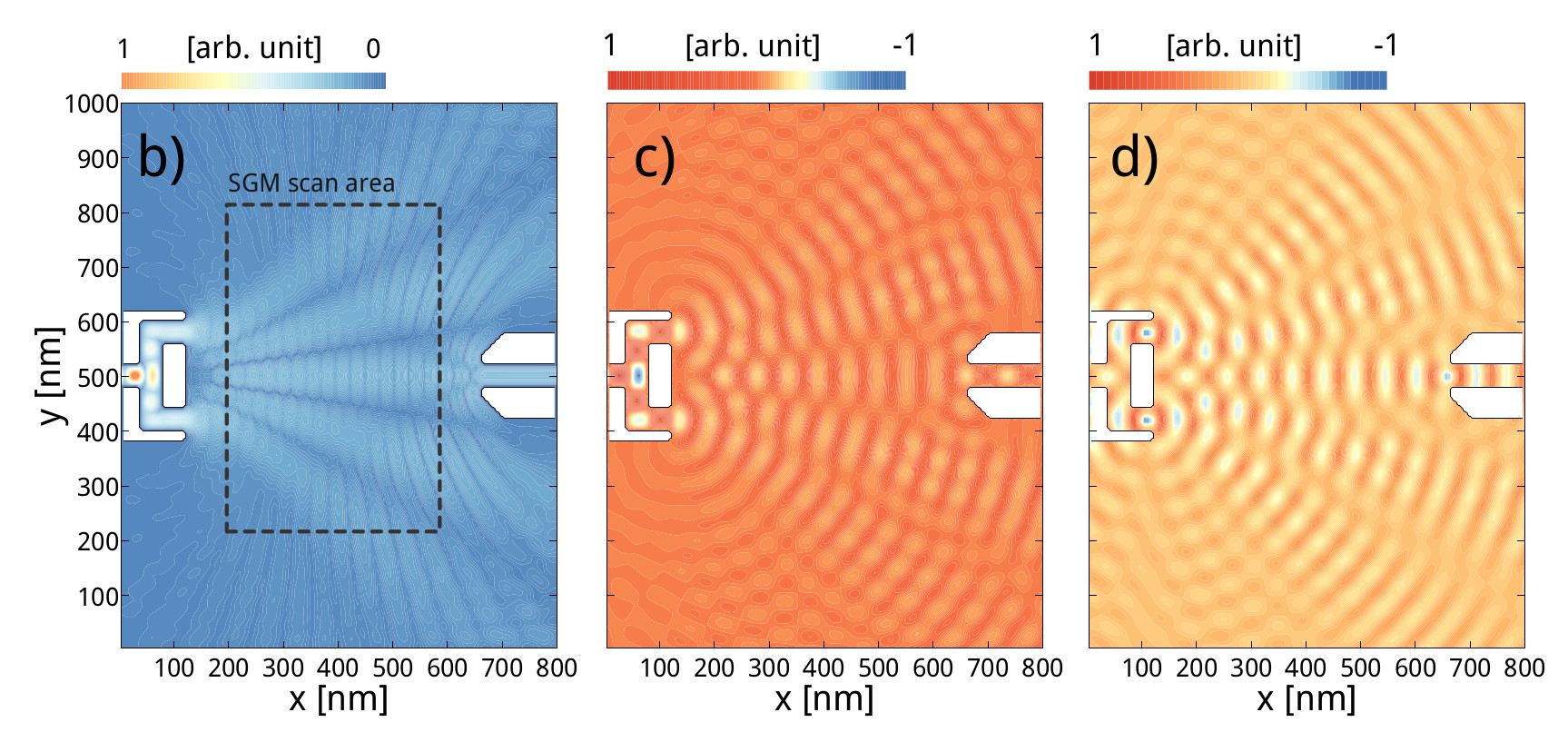} \tabularnewline
\end{tabular}
\par\end{centering}

\caption{a) Sketch of the system considered in this paper with the double-slit
system -- source channel -- QPC1 of width $w_{\mathrm{qpc1}}$ that
splits the incoming electron wave function into two beams. The width of the QPC1 slits equal to the width of the input channel is kept at $w_\mathrm{qpc1}=40$ nm
all along this work. The slits
of QPC1 in this Figure are separated a distance of $d_{\mathrm{qpc1}}=160$ nm. In the rest of the paper $d_{\mathrm{qpc1}}=400$ nm is mainly used.
The outgoing
current is gathered by the drain channel -- QPC2 at right of
width $w_{\mathrm{qpc2}}$ to determine source-drain conductance $G$. Additionally the current between the source
and the rest of the system excluding the drain detector
is measured for evaluation $G_{\mathrm{rest}}$ leakage conductance. The sample is assumed large and the blue dashed line
indicates the ends of the computational box where transparent boundary
conditions are applied. The distance between the
computational box and the edge of the sample
($l_{d}$) is assumed much bigger than the coherence length
 $l_{d}\gg\lambda_{\mathrm{coherence}}$.  The blue rectangle, between the left and right
QPCs, shows the area of calculated SGM conductance maps.
The distance between QPC1 and QPC2 is 600 nm. b) The absolute
value of the scattering wave function $|\psi|$. c) Real part of $\psi$.
d) Real part of the wave function calculated from equation (\ref{eqfaleKuliste})
as superposition of three point sources in the centers of the slits.}

\label{figModel}
\end{figure*}

We consider an experimental situation that is depicted in Fig. \ref{figModel}(a).
The current is fed by the source contact to the channel that is filtered
by the split quantum point contact QPC1. The second QPC (QPC2) serves
as an entrance to the drain contact at the right side of the figure
that will be used for electron detection as in Refs. \cite{crook,aidala,skipping}
or more recently in Ref. \cite{khatua2014}. The effective width of
the source channel $w_{qpc1}$ and the width of both slits at the
QPC1 side is taken equal 40 nm, which for the considered Fermi energy
transmits the current in the lowest subband only. According to the
previous paper \cite{Kolasinski2014} the lowest subband transport
on the input part is necessary for the Young interference to be observed.
For a larger number of incident subbands the conductance map for the
double slit system is a  simple sum of maps for separate QPCs \cite{Kolasinski2014}
since the double-slit interference disappears in the Landauer summation over the incident subbands.
The electron leaving QPC1 enters a finite but large sample filled
by 2DEG. By large we mean that the $l_{d}$ distance between the region
of interest for the SGM and the edges of the sample is larger than
the coherence length, so that the interference with the edges can
be neglected. The source-drain voltage is considered low enough for
the linear transport conditions to occur. The current passing to the
drain is measured in order to evaluate the $G$ conductance. The sample
is grounded by a large reflectionless contact and this current is
measured to evaluate the leakage conductance $G_{rest}$. From conservation
of the current and for $M_{in}$ subband in the input channel (we consider
mostly $M_{in}=1$) we have
\begin{equation}
\frac{2e^{2}}{h}M_{in}=R+G+G_{\mathrm{rest}},\label{eqM}
\end{equation}
where $R=\frac{2e^{2}}{h}P_{bs}$, where $P_{bs}$ is the backscattering
probability. Thus, $R$ can be determined when $G$ and $G_{rest}$
are measured. In the following we discuss numerical results for $R$
and $G$.

In the calculations we focus on the region marked by the dashed lines
in Fig. \ref{figModel}(a). We assume that the transport is coherent within
the computational box.
At the dashed lines we apply transparent boundary conditions,
so that the region of interest is effectively open, in contrast to
systems with a pair of QPCs used as source and drain for a closed
stadium that are studied by SGM in e.g. Refs. \cite{kozikov,kozikov2}.
In this paper we set the distance between the input slits $d_{\mathrm{qpc1}}=400$nm,
which is large enough to distinguish trajectories of the electron
arriving to QPC2 from one of the input slits. In Fig. \ref{figModel} a smaller value of  $d_{\mathrm{qpc1}}=160$nm is used
for illustration. The total computational
box covers the region where the transparent boundary conditions are applied
of size 800nm$\times$1000nm and the region where the scans by the
SGM are taken is significantly smaller: 400nm $\times$ 600nm -- see
Fig. \ref{figModel}(a).

In order to simulate the propagation of the Fermi level electrons
within the region inside the computational box (dashed lines in Fig.
\ref{figModel}(a)) we consider coherent transport as described by
the effective-mass Schrödinger equation,
\begin{equation}
\left\{ -\frac{\hbar^{2}}{2m_{\mathrm{eff}}}\nabla^{2}+V_{\mathrm{tip}}(x,y)\right\} \psi(x,y)=E_{\mathrm{F}}\psi(x,y),\label{eqschrodinger}
\end{equation}
where $m_{\mathrm{eff}}=0.067m_{\mathrm{0}}$ is GaAs electron effective
mass, $E_{\mathrm{F}}$ is the Fermi level energy, and potential
\begin{equation}
V_{\mathrm{tip}}(x,y)=\frac{U_{\mathrm{tip}}}{1+\left[\left(x-x_{\mathrm{tip}}\right)^{2}+\left(y-y_{\mathrm{tip}}\right)^{2}\right]/d_{\mathrm{tip}}^{2}}\label{eqlorentz}
\end{equation}
describes the effective perturbation induced by the AFM tip with amplitude
$U_{\mathrm{tip}}=15$ meV and width $d_{\mathrm{tip}}$ , localized
above point $x_{tip},y_{tip}$. This type of effective Lorentzian-shaped
perturbation, which results from the screening of the charge on the
tip by the 2DEG electron gas located under the sample surface, was
obtained previously in our self-consistent Schrödinger-Poisson calculations
\cite{kolasinski2013,szafran}. The width of the potential is of the
order of the distance between the tip and the 2DEG \cite{szafran}.
In the SGM experiments the 2DEG is buried at least 25 nm below the
surface \cite{atleast}, and the minimal distance of the tip to the
surface applied in SGM is 20 nm \cite{sgmr1}, hence a minimal realistic
value of $d_{tip}$ is about 50 nm. In this paper we consider two
widths of the tip: a small one $d_{tip}=10$ nm which is useful for
the initial discussion since it sets the precision of the determination
of electron paths, and a  realistic one -- $d_{tip}=50$ nm.

For simplicity in order to define the contacts we assume hard-wall
boundary conditions on QPC1 and QPC2. In experimental setups the QPCs are usually defined
electrostatically by potential applied between the gates so that the
QPC potential has  a saddle point profile. Nevertheless, we
do not discuss conductance quantization as function of the QPC width
and for a fixed number of transmitting subbands the hard-wall potential
gives qualitatively similar results to a saddle point potential which
was applied in a previous work \cite{Nowak2014}. The transparent
boundary conditions at the blue dashed line of Fig. \ref{figModel}(a))
were introduced by the method described in Ref. \cite{Nowak2014}.

We work within the Landauer approach for a zero temperature in the
finite difference implementation \cite{Kolasinski2014QHI} of the
quantum transmitting boundary method \cite{Lent90,Lent94}. Within
the input channel -- before the splitting ending with the two QPC1 slits, the wave function at the Fermi level is given by
the superposition of incoming and backscattered transverse modes
\begin{eqnarray}
\psi_{\mathrm{qpc1}}(x,y) & = & \sum_{k=1}^{M_{\mathrm{qpc1}}}\left\{ a_{k}e^{ikx}\chi_{k}^{\mathrm{qpc1}}(y)+b_{k}e^{-ikx}\chi_{-k}^{\mathrm{qpc1}}(y)\right\} \nonumber \\
 & + & \sum_{k=M_{\mathrm{qpc1}}+1}^{+\infty}b_{k}e^{kx}\chi_{k}^{\mathrm{qpc1}}(y),\label{eqsolQPC1}
\end{eqnarray}
where the last term corresponds to the summation over the evanescent
modes \cite{Lent90,Lent94,Bagwell1990} and $M_{\mathrm{qpc1}}$ is
the number of current propagating transverse modes $\chi_{k}^{\mathrm{qpc1}}$
in QPC1. We consider a single subband in the QPC1 channel and an arbitrary
number of subbands in the QPC2 channel. The summation runs over the
Fermi level wave vectors $k$ at subsequent lateral subbands.
The transport from the input channel to the two-slits is non-adiabatic with a pronounced backscattering.
Note, that the central island [Fig. 1(a)] used as a beam splitter in the presence of a strong external magnetic field is likely to form a quantum Hall
interferometer as discussed in Refs. \cite{ensliny,martinsy}

For QPC2 we assume the boundary conditions of form
\begin{eqnarray}
\psi_{\mathrm{qpc2}}(x,y) & = & \sum_{k=1}^{M_{\mathrm{qpc2}}}d_{k}e^{ikx}\chi_{k}^{\mathrm{\mathrm{qpc2}}}(y)+\nonumber \\
 &  & \sum_{k=M_{\mathrm{qpc2}}+1}^{+\infty}d_{k}e^{-kx}\chi_{k}^{\mathrm{\mathrm{qpc2}}}(y),\label{eqsolQPC2}
\end{eqnarray}
with $M_{\mathrm{qpc2}}$ being the number of conducting transverse modes in
QPC2. The transverse modes $\chi_{k}$ for both QPCs were calculated
using the method of Ref. \cite{Zwierzycki2008}. The scattering amplitudes
$b_{k},\, d_{k}$ and the amplitudes of the incoming modes $a_{k}$
of Eq. (\ref{eqsolQPC1}) once established are used to calculate the
transmission probabilities. Throughout this paper we keep $E_{\mathrm{F}}=8$meV
which gives the value of $\lambda_{\mathrm{F}}=2\pi/k_{\mathrm{F}}=2\pi\hbar/\sqrt{2m_{\mathrm{eff}}E_{\mathrm{F}}}\approx53$nm.
We choose discretization grid spacing $\Delta x=4$nm, which is small
compared to $\lambda_{\mathrm{F}}$.

After solution of Eq. \ref{eqschrodinger} for each incoming mode
the source-drain conductance is evaluated from the Landauer formula
\begin{equation}
G=G_{0}\sum_{i=1}^{M_{\mathrm{qpc1}}}T_{i},\label{eqG}
\end{equation}
where $T_{i}$ is the transmission probability of the $i$-th mode
incoming from the QPC1 to the QPC2 and $G_{0}=2e^{2}/h$. We refer
to the sum of backscattering probabilities as the ''resistance'',
which is given by the formula
\begin{equation}
R=G_{0}\sum_{i=1}^{M_{\mathrm{qpc1}}}R_{i},\label{eqR}
\end{equation}
where $R_{i}$ is the backscattering probability of $i$-th incoming
mode to the QPC1.

We discuss below the maps of backscattering probability (SGM-R) and
conductance maps (SGM-G) for current flow between QPC1 and QPC2.
In order to calculate the SGM-R/G images we evaluated the ''resistance''
R and conductance G by scanning the system with the tip potential given
by Eq. \eqref{eqlorentz} inside the rectangle shown in Fig. \ref{figModel}(b).
The spacing between two subsequent positions of the tip was $4$nm,
thus in order to obtain the SGM image, presented later in paper, for
the scan area 400nm $\times$ 600nm one has to solve the scattering
problem 15000 times.

\section{Results and discussion}
\subsection{In the absence of the tip: Young interference and interference due to QPC2 detector}
We begin the discussion by setting equal widths of both input and
output channels within QPCs, $w_{\mathrm{qpc1}}=w_{\mathrm{qpc2}}=40$nm.
In both the input and the output channels we have a single subband transport
for the applied Fermi energy of $8$ meV. According to our previous
paper \cite{Kolasinski2014} the single subband transport on the input
part allows for resolution of the Young interference pattern by the
SGM technique.

The amplitude of the scattering wave function in the absence of the
tip is shown in Fig. \ref{figModel}(b). 
Besides the well-resolved Young interference pattern with five more
or less straight lines of constructive and destructive interference
for the waves passing through both the slits [see Fig. \ref{inter}(c)]
one also notices additional vertical interference fringes. This feature
is caused by the presence of the output QPC2 which reflects the incoming
wave function back to the double slit device.

The role of QPC2 in the scattering is more clearly visible in the
real part of the wave function displayed as Fig.\ref{figModel}(c).
We found that a very similar wave function can be reproduced by a
superposition of three point sources positioned at the center of each
slit of the system following the Huygens\textendash{}Fresnel principle
for the circular waves propagating from the input slits and QPC2 --
the latter as due to backscattering of the incoming wave \cite{Gorini2013}.
In the limit of thin slit the Fermi level wave function at position
$\boldsymbol{r}$ traveling from the slit located at position $\boldsymbol{r}_{\mathrm{source}}$
can be well described by the angle-modulated Henkel function
\begin{eqnarray}
 &  & \psi_{\mathrm{Henkel}}(r,r_{\mathrm{source}},\alpha_{\mathrm{source}})=\nonumber \\
 &  & \alpha_{\mathrm{source}}\cos(\theta)\frac{e^{ik_{\mathrm{F}}|{\bf r}-{\bf r}_{\mathrm{source}}|}}{\sqrt{k_{\mathrm{F}}|{\bf r}-{\bf r}_{\mathrm{source}}|}},\label{eqHenkelPsi}
\end{eqnarray}
where $\theta$ is angle between ${\bf r}-{\bf r}_{\mathrm{source}}$
vector and the $x$ axis, i.e. $\cos(\theta)={|x-x_{\mathrm{source}|}}/{r}$
and $\alpha_{\mathrm{source}}$ is a scattering amplitude. For the superposition
of the three point sources one has
\begin{eqnarray}
\psi(r) & \approx & \psi_{\mathrm{Henkel}}(r,r_{1},1)+\psi_{\mathrm{Henkel}}(r,r_{2},1)+\nonumber \\
 &  & \psi_{\mathrm{Henkel}}(r,r_{3},\alpha_{\mathrm{reflection}}),\label{eqfaleKuliste}
\end{eqnarray}
where $r_{1}=\left(110\mathrm{nm},-80\mathrm{nm}\right)$, $r_{2}=\left(110\mathrm{nm},+80\mathrm{nm}\right)$
and $r_{3}=\left(670\mathrm{nm},0\mathrm{nm}\right)$ are the positions
of slits in the system of Fig. \ref{figModel}(b). We set $\alpha_{\mathrm{reflection}}=\alpha_{\mathrm{source}}/2$,
for which we get the best agreement with the exact solution obtained
from Eq. (\ref{eqschrodinger}). The real part of the wave function
obtained from Eq.(\ref{eqfaleKuliste}) is plotted in Fig. \ref{figModel}(d).
Note that the fitted value of $\alpha_{\mathrm{reflection}}$ is quite
large, which suggests that the backscattering from QPC2 will have
a substantial influence on the SGM images discussed below.

The Young interference [Fig. \ref{inter}(d)] and the interference
due to backscattering by the QPC2 detector [Fig. \ref{figModel}(b-c)]
are present already without the tip.
The Young interference involves superposition of waves passing through each of the slits with the modulation of the scattering density given by
\begin{equation}
\rho = \cos(k_F(r_1'-r_2')).\label{yo}
\end{equation}

\subsection{Interference mechanisms due to the tip}

\begin{figure*}[htbp]
\begin{centering}
\includegraphics[width=0.8\paperwidth]{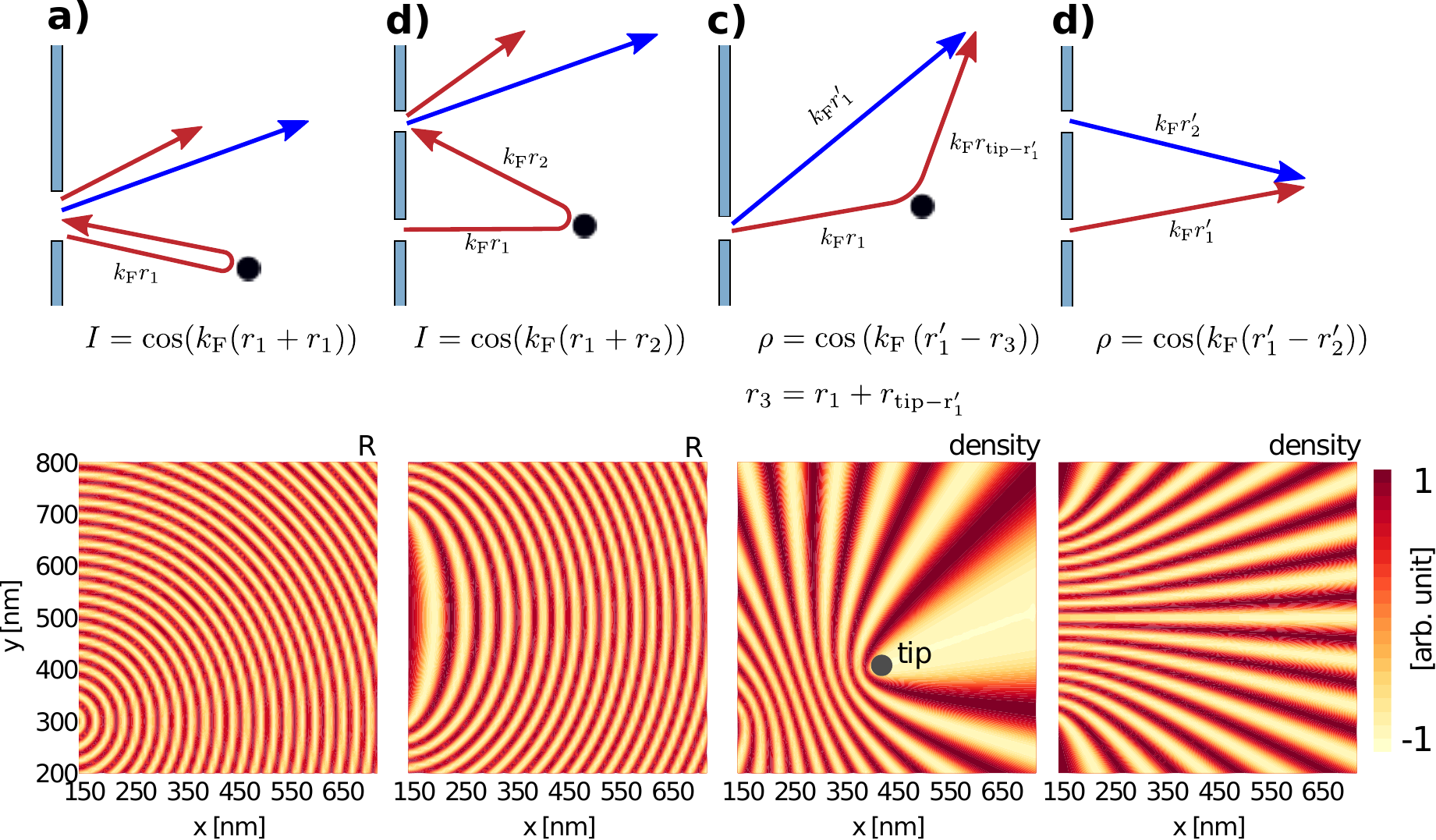}
\caption{Schematic presentation of the discussed interference processes.
Upper panels show the interfering paths (blue and red), and lower panels plot
the cosine of the phase difference as a function of the tip position (a,b) for
the SGM-R maps (a,b) or -- in the absence of the tip (c,d) -- for the scattering density.
The following interference mechanisms are discussed below:
 a) due to the scattering of the incoming wave on the tip
potential wave function may be reflected backward the input slit.
Such process will lead to circular fringes around the QPC1
slit entrance in the SGM-R images -- see the lower panel and Eq. \ref{intera}.
b) tip induced double-slit interference
when the electron wave is backscattered by the tip to the other slit that leads to elliptic fringes visible
in SGM-R images -- see the lower panel and Eq. (\ref{eq:model2})
c) tip-induced interference
of unperturbed wave with the wave reflected from the potential tip
(this process leads to the lateral fringe pattern in SGM-G images
and upon additional backscattering by the detector to the lateral fringe
pattern in SGM-R images) -- see Eq. (\ref{eq:model3}), d) Young-type double slit interference
of two coherent circular waves (characteristic beams visible in SGM-R images) -- see Eq. (\ref{yo}).}
\label{inter}
\par\end{centering}

\centering{}
\end{figure*}

The backscattering by the tip introduces
additional interference effects which will be discussed below: with
the electrons reflected to the source QPC [Fig. \ref{inter}(a)],
the tip-induced double slit interference [Fig.
\ref{inter}(b)] and the interference of the direct wave with the scattered one [Fig.
\ref{inter}(c)].  The conductance map fringes due to the scattering
of the type given by Fig. \ref{inter}(a) is well known and has been discussed
in a number of papers \cite{int1,int2,int3,int4,Jalabert2010,Gorini2013,Brun2014,kozikov2013}.
The backscattering ($R$) is enhanced when the phase shift of the wave going to the tip and back [red arrows in Fig. \ref{inter}(a)]
produces a constructive interference with the incident wave [blue arrow in Fig. \ref{inter}(a)] at the input slit.
The $R$ signal is then proportional to
\begin{equation}
I=\cos(k_{\mathrm{F}}(r_{1}+r_{1})),\label{intera}
\end{equation}
where in the argument $k_F \times 2 r_{1}$ is the phase acquired by the electron wave function
on its way from a slit of QPC1 to the tip and back from the tip to the same slit.
The interference of Fig. \ref{inter}(a)
leads to the angle-independent modulation of the $R$ maps
which oscillate as a function of distance from the slit with
the period of half the Fermi wavelength $\lambda_{\mathrm{F}}/2$
\cite{Jalabert2010,Gorini2013}.

The tip induces a double-slit interference of the wave passing through one of the slits of QPC1 and scattered by the tip
with the wave incident from the other  QPC1 slit [Fig. \ref{inter}(b)]. Then, an enhanced
backscattering can be expected when the interference of the incident and returning path is positive,
or the $R$ signal will be proportional to
\begin{equation}
I=\cos\left(k_{\mathrm{F}}\left(r_{1}+r_{2}\right)\right).\label{eq:model2}
\end{equation}
The result of formula (\ref{eq:model2}) is plotted in Fig. \ref{figsgmr}(g),
with a good agreement with the oscillations found in the results of
the scattering problem of Fig. \ref{figsgmr}(e-f).
The double slit interference according to
mechanism of Fig. \ref{inter}(b) and the resulting SGM effects were
never discussed before. The equations (\ref{intera} and \ref{eq:model2}) produce the
same periodicity of $\lambda_{\mathrm{F}}/2$ at a large distance from QPC1.

The tip induces also interference of the waves incident from one of the QPCs
and backscattered by the tip -- as depicted in Fig. \ref{inter}(c).
The modulation of the scattering density for a fixed tip position is given by
\begin{equation}
\rho=\cos\left(k_\mathrm{F} (r'_1-r_3)\right)  \label{eq:model3}
\end{equation}
with $r_3=r_1+r_{\mathrm{tip}-r_1'}$,
and $r_{\mathrm{tip}-r_1'}$ is the distance between the tip
and ${r_1'}$. The results of Eq. \ref{eq:model3} are displayed in the lower panel of Fig. \ref{inter}(c)
for a given tip position. This type of interference leads to lateral fringe patterns in SGM-G images to be discussed below.

\subsection{Resistance maps }

\begin{figure*}[htbp]
\centering{}\includegraphics[width=0.7\paperwidth]{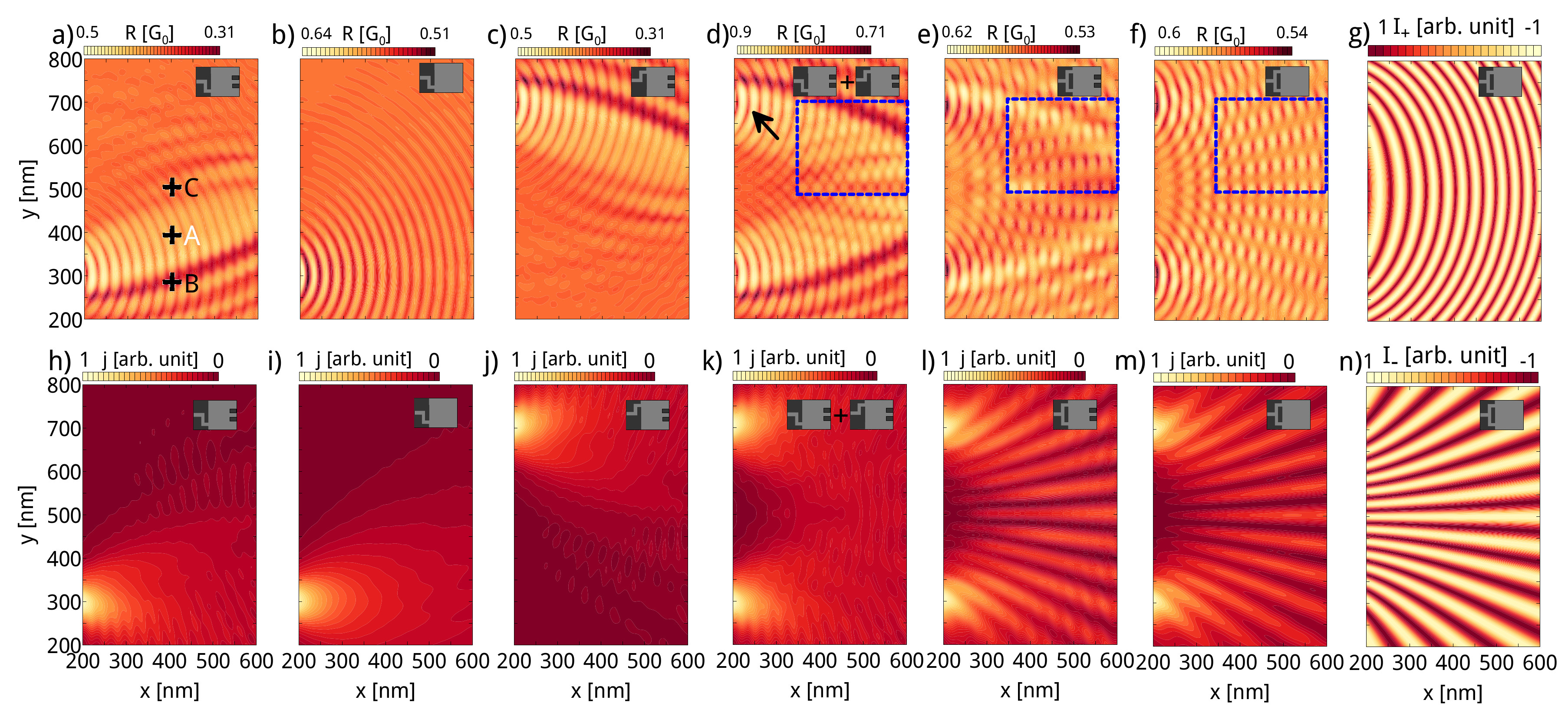} \caption{First row -- (a-f) -- shows the SGM-R images obtained by scanning
the rectangle area indicated in Fig. \ref{figModel}(b) by AFM tip
described by potential Eq.(\ref{eqlorentz}). The insets in each figure
show schematically the device for which the image was calculated.
Second row -- (h-m) -- shows the {amplitude of} probability current
distribution calculated for system without external potential of the
tip. bottom slit open (a,h) and (b,i), upper slit open (c,j), a sum
(a,c) and (h,j) is given in (d) and (k), respectively. (e,l) and (f,m)
show the results for both input slits open. In (e,l) QPC2 is present, and in (f,m) it is removed. (g,n) are the images obtained
from Eq. \eqref{yo} and Eq. \eqref{eq:model2}. The results were obtained for $U_{\mathrm{tip}}=15$ meV and $d_\mathrm{tip}=10$nm. \label{figsgmr}}
\end{figure*}

In Fig. \ref{figsgmr}(a-f) we show the SGM maps of resistance (SGM-R)
as functions of the position of the AFM tip. The area of the scan
is shown by the blue rectangle in Fig. \ref{figModel}(a,b). We consider
a single or both QPC1 slits open as illustrated schematically in the insets
in the top-right corner. For each system we plotted in the second
row [Fig. \ref{figsgmr}(h-m)] the corresponding probability current
distribution. In Fig. \ref{figsgmr}(a,b) we show the results of SGM-R
images obtained for a single (lower) input slit open. In the absence
of QPC2 [Fig. \ref{figsgmr}(b)] we observe  circular fringes
in the SGM map due to the interference of type given by Fig. \ref{inter}(a)
between the wave incoming from the slit and the wave backscattered
by the AFM tip [see the arrow in Fig. 3(d), Fig. \ref{inter}(a), and Eq. \ref{intera}].

\begin{figure*}[htbp]
\begin{centering}
\includegraphics[width=0.7\paperwidth]{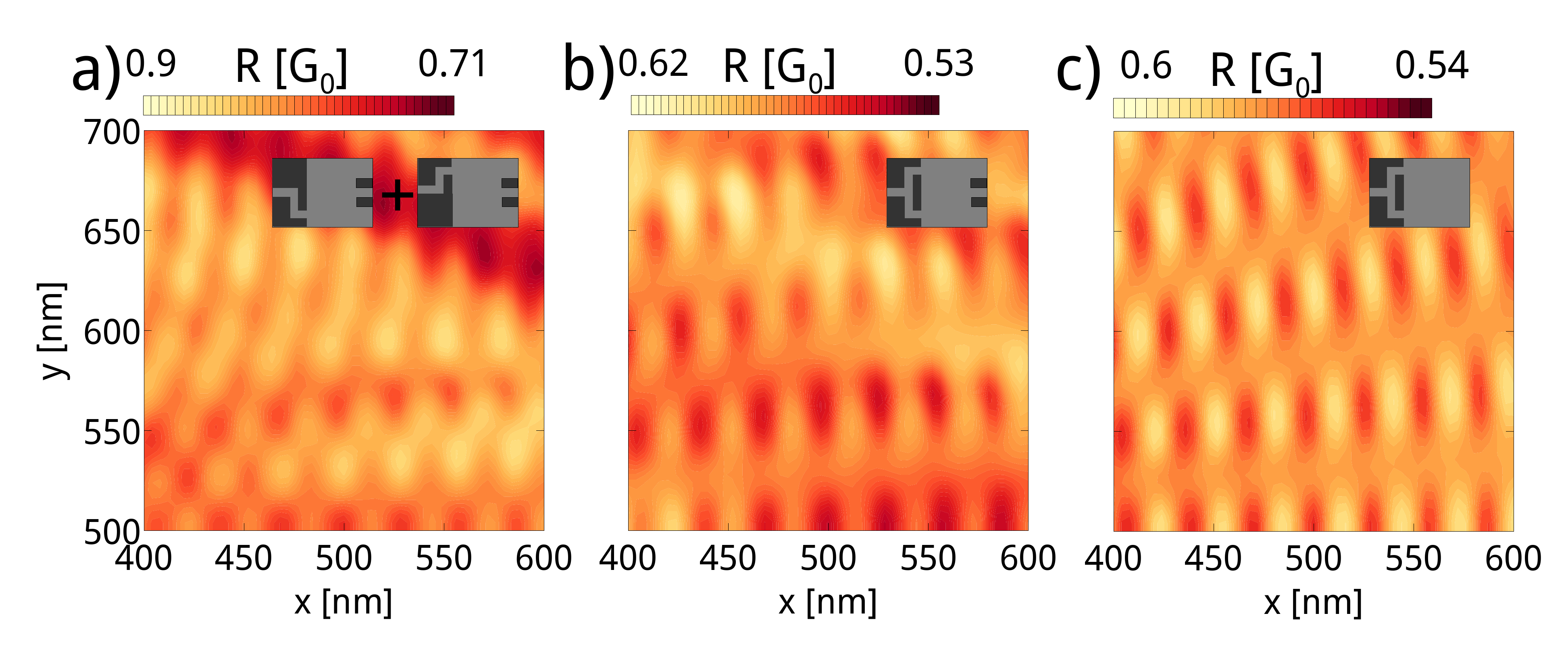} \caption{The zooms of Figures \ref{figsgmr}(d-f) denoted by blue rectangles.}
\label{figzumy}
\par\end{centering}

\centering{}
\end{figure*}

Figure \ref{figsgmr}(d) presents the sum of SGM-R maps of Fig. \ref{figsgmr}(a)
and (c) obtained for a single slit open. The sum is quite different from the image obtained for the
system where both the QPC1 slits are open (see Fig. \ref{figsgmr}(e)),
which is a signature of the double-slit interference effects, and
involve both the Young interference [Fig. \ref{inter}(d)] and
the tip-induced interference [Fig. \ref{inter}(b)]. 
The Young interference is more clearly resolved in the absence of QPC2 [Fig.
\ref{figsgmr}(f)], although it is also present when the QPC2 detector
is a part of the setup [Fig. \ref{figsgmr}(e)]. In order to demonstrate
this closer we extracted the enlarged fragments of Fig. \ref{figsgmr}(e,f)
in Fig. \ref{figzumy}(b,c). For a simple sum of SGM images for separate
slits [Fig. \ref{figsgmr}(d)] instead the elliptic fringes or Young pattern we observe a checkerboard pattern [Fig. \ref{figsgmr}(d)
and Fig. \ref{figzumy}(a)]. This checkerboard pattern should be
observed in the experiments for a large number of incident subbands,
i.e. at higher Fermi energy or wider input channel as discussed in
Ref. \cite{Kolasinski2014}.

In Figs. \ref{figsgmr}(h-m) we plotted the current density within
the system in the absence of the tip. Note, for the double slit systems
[Fig. \ref{figsgmr}(l,m)] that the current distribution is very
similar with or without QPC2. The deformation of the Young interference
for the system with QPC2 is due to the lateral interference pattern
involving the paths marked in Fig. \ref{inter}(b) which are introduced by QPC2.

The Young interference pattern
given by Eq. (\ref{yo}) and calculated for the Fermi wave vector
is displayed in Fig. \ref{figsgmr}(n) with a good agreement with
the probability current distribution of Fig. \ref{figsgmr}(l,n) and
the features of the SGM map of Fig. \ref{figsgmr}(f). A resemblance
to Fig. \ref{figsgmr}(e) -- for QPC2 present -- can also be spotted,
although the lines of flat $R$ in Fig. \ref{figsgmr}(e) are curved.
The curvature as well as the presence of the lateral fringes can
be reduced by placing the QPC2 further from QPC1, thus reducing the
backscattering by QPC2 but at the expense of the reduced $G$ map
contrast (see below).

\subsection{Source-drain conductance maps}

\begin{figure*}[htbp]
\begin{centering}
\includegraphics[width=0.6\paperwidth]{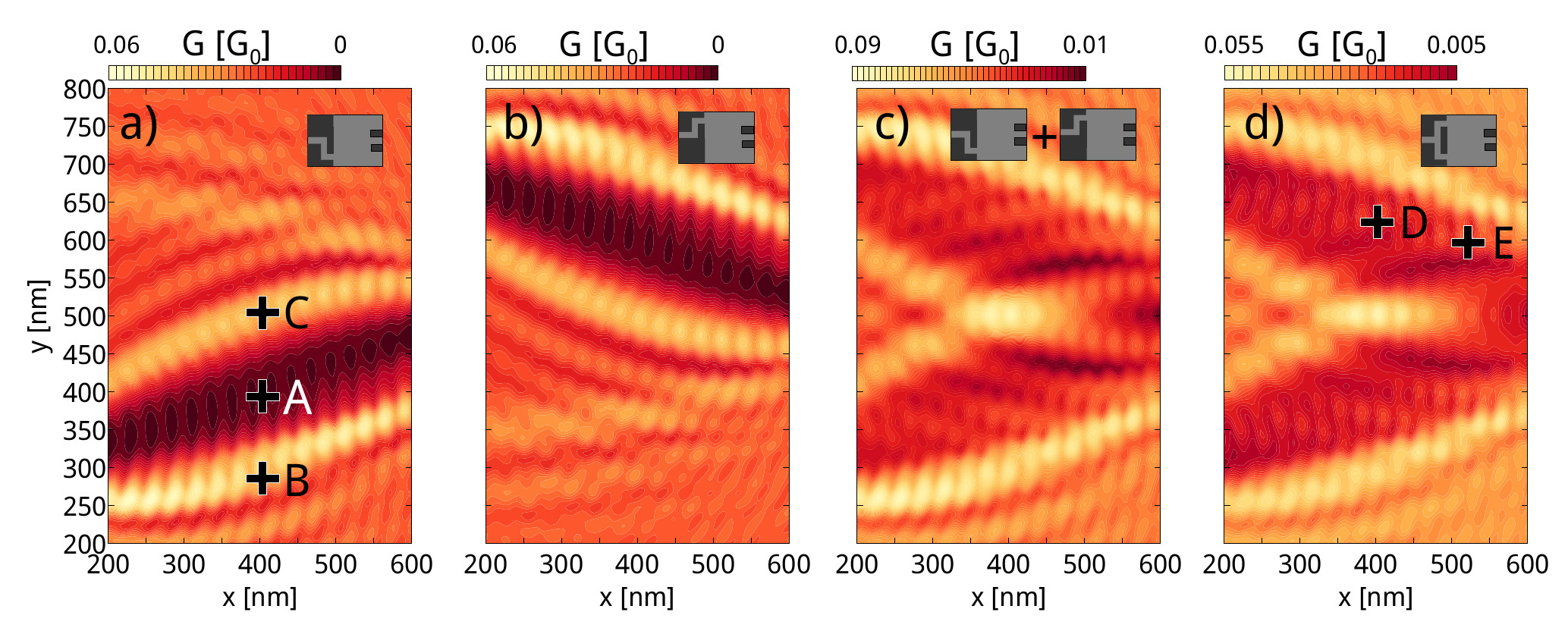}
\par\end{centering}

\caption{The SGM conductance maps obtained for only the bottom (a) or upper
(b) slit open; (c) sum of (a) and (b); (d) map for both slits open.
Points A-E marked by the crosses show the tip positions for which
we plot the probability current distribution in Fig.\ref{figprady}. }

\label{figsgmg}
\end{figure*}

The SGM maps of source-drain conductance (SGM-G) of Fig. \ref{figsgmg}
exhibit a valley of minimal values  along the shortest path from
the source channel to the drain. The pattern of the image obtained
for both slits open [Fig. \ref{figsgmg}(d)] is exactly the same
as the one given by the sum of images Fig. \ref{figsgmg}(a) and (b)
(see Fig. \ref{figsgmg}(c)). This shows that in SGM-G images the
double-slit interference is absent, which may be quite surprising,
given that the SGM-R image clearly resolved the interference.

\begin{figure*}[htbp]
\begin{centering}
\includegraphics[width=0.8\paperwidth]{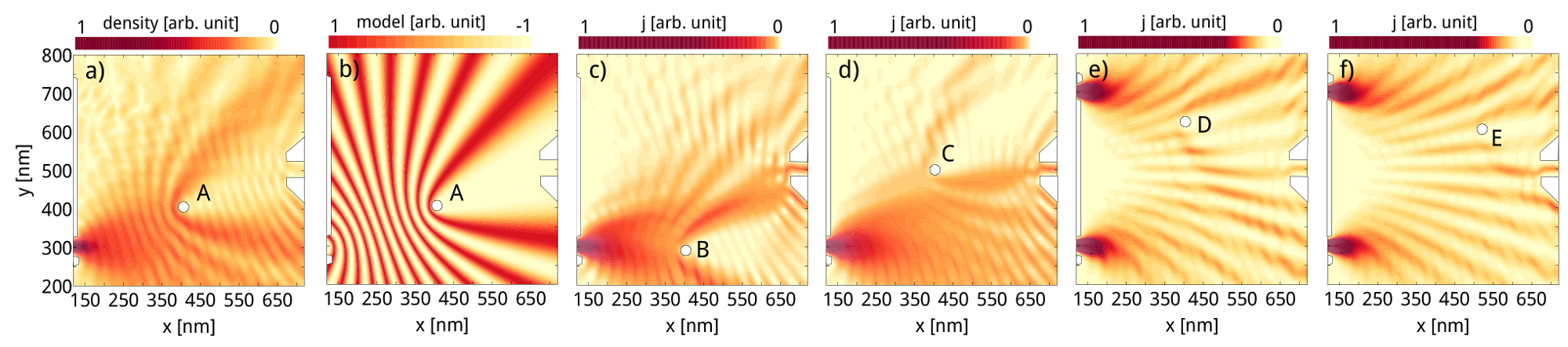}
\par\end{centering}

\caption{The probability current obtained for five different positions of the
AFM tip denoted by white dots and marked in Fig. \ref{figsgmg}. (a,c,d)
Result for bottom slit open obtained for points A, B and C in Fig.
\ref{figsgmg}(a). b) result obtained from simple model given by Eq.
\eqref{eq:model3}. e-f) Results for both slits open obtained for
points D and E in Fig.\ref{figsgmg}(d), respectively.}

\label{figprady}
\end{figure*}

In Fig. \ref{figsgmg}(a)
a lateral fringe pattern is observed along the classical trajectory.
This valley of minimal $G$ values coincides with the line of maximal backscattering $R$
observed in Fig. 3(a).  Figure \ref{figprady}(a) shows that the tip separates the electron
wave into beams, which arise from the interference
of waves incoming from QPC1 and scattered by the tip -- as illustrated
by Fig. \ref{inter}(c). The variation of the electron density can
be approximated by Eq. \ref{eq:model3}.
The results of formula (\ref{eq:model3}) are plotted
in Fig. \ref{figprady}(b) with a very good agreement with the numerical
result of Fig. \ref{figprady}(a), which however contains also the
fringes due to backscattering by QPC2.
 Figure \ref{figprady}(a)
corresponds to the position of the tip above {the} point A and explains the valley
of low conductance
in the SGM-G image (e.g. see Fig. \ref{figsgmg}(a)), since QPC2
is in the shadow cast by the tip.
Figures \ref{figprady}(c-d) correspond to tip above points B and
C for a single slit and explain the increase in the conductance around
the classical path, which is well visible in Fig. \ref{figsgmg}(a)
and (b). If we move the tip to positions B or C (see Fig. \ref{figprady}(b)
and (c)) the electron wave is still separated into two main beams with
one of these beams reaching QPC2, which produces the characteristic
envelope of high conductance around the ``classical path''. The
minima of the lateral pattern of Fig. \ref{figsgmg} appear when
a nodal line of the interference passes to the QPC2 detector.

 The points $A,B,C$ in Fig. \ref{figsgmr}(a)
are the same as in Fig. \ref{figsgmg}(a) and correspond to plots
Fig. \ref{figprady}(a-d). Note that point $B$ corresponds to a minimum
of $R$ and maximum of $G$, while point $C$ corresponds to a maximum
of both quantities. The lateral pattern in the $R$ map appears only
in presence of QPC2 detector [cf. Fig. \ref{figsgmr}(a) and Fig.
\ref{figsgmr}(b)], and is a result of scattering first by the tip
and next by QPC2. The effect of the interference of type Fig. \ref{inter}(b) depends on
the 1) position of the tip with respect to QPC2 - QCP1 -- return path
for the backscattered electrons and 2) the incidence angle of the
waves deflected by the tip on the edges of the QPC2 electron.

Let us now consider both slits open with the AFM tip at position where
the electron flow from the upper slit is totally blocked (see Fig.
\ref{figprady}(e)). The current flux from the lower slit will be
the only one reaching the QPC2. The tip in this manner turns off one
of the slits of the source channel. Using the symmetry arguments for the considered
device we will get the same conclusion with second slit blocked and
the first transmitting current. Such argumentation can explain why
at certain points in the obtained SGM-G maps there is no visible interference.
Nevertheless, there are points where the potential of the tip does
not totally block the current from any slit, so an interference could
be expected. Such a case is presented in Fig. \ref{figprady}(f),
with the tip located at a position where the current distribution
is almost the same as in the unperturbed system (for comparison see
Fig. \ref{figsgmr}(l)). We conclude that the double-slit interference
is clearly visible in the current distribution but not in the SGM-G
images.

\begin{figure}[htbp]
\begin{centering}
\includegraphics[width=0.4\paperwidth]{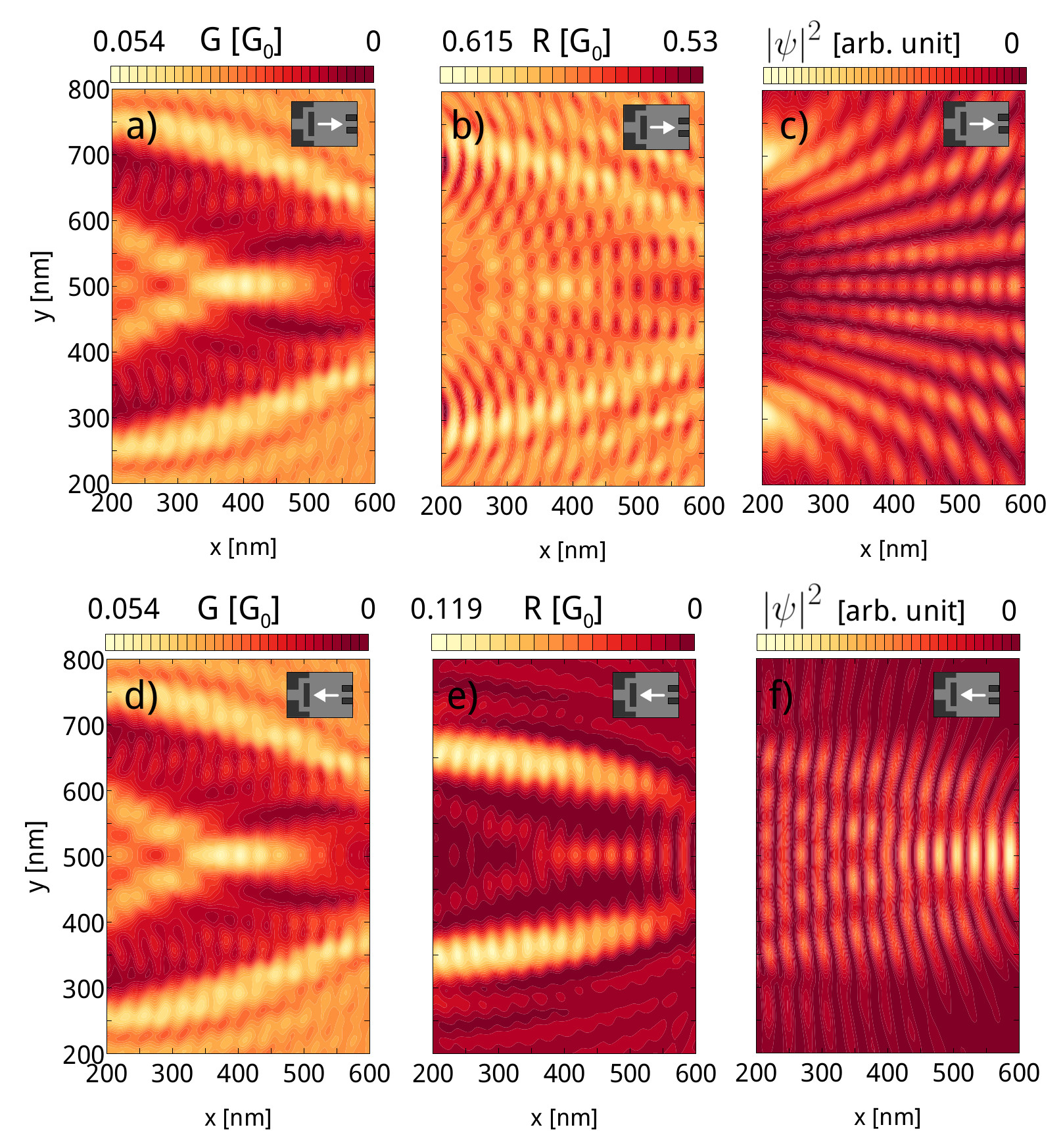}
\par\end{centering}

\caption{First row a) The SGM-G image, b) SGM-R image c) and the electron density
$|\psi|^{2}$ obtained for current flow to the right direction in
the absence of the tip. Second row d-f) The same as in the first row
but with current flow in the opposite direction. The results were obtained for $U_{\mathrm{tip}}=15$ meV and $d_\mathrm{tip}=10$nm. }

\label{figodtylu}
\end{figure}

The lack of Young interference in the source-drain conductance  images can be
explained using a reversed bias and the current flowing in the reverse
direction, i.e. from QPC2 to QPC1. The upper row of Fig. \ref{figodtylu}
shows the SGM-G, SGM-R and probability density for the current flowing
from the double slit to QPC2, while the lower presents the quantities
for the opposite current direction. We can see that the results for
SGM-G are exactly identical for both current directions [Fig. \ref{figodtylu}(a,d)]
in spite of the fact that the electron scattering is very different
in both setups with a clear Young interference in Fig. \ref{figodtylu}(c)
with no similar counterpart in Fig. \ref{figodtylu}(f).

The SGM-G images for both the cases [Fig. \ref{figodtylu}(a,d)]
is bound to be identical due to the Onsager microreversibility relation
for the single subband transport $T_{QPC2,QPC1}=T_{QPC1,QPC2}$, which
in terms of the Landauer approach implies no current flow for zero
bias. For the electron incident from the right there is no reason
to expect a presence of the Young interference, which is indeed missing
in Fig. \ref{figodtylu}(f). The absence of the double
slit interference in SGM-G follows from this observation and the microreversibility
relation. Note, that SGM-R image for the reversed current orientation
[Fig. \ref{figodtylu}(e)] clearly indicates the classical paths
that the electron can follow from QPC2 to QPC1.
Note, that QPC2 is further 100 nm to the right of the end of the figure and that two bright beams are emitted from QPC2 which is only 40 nm wide.

\begin{figure*}[htbp]
\begin{centering}
\includegraphics[width=0.8\paperwidth]{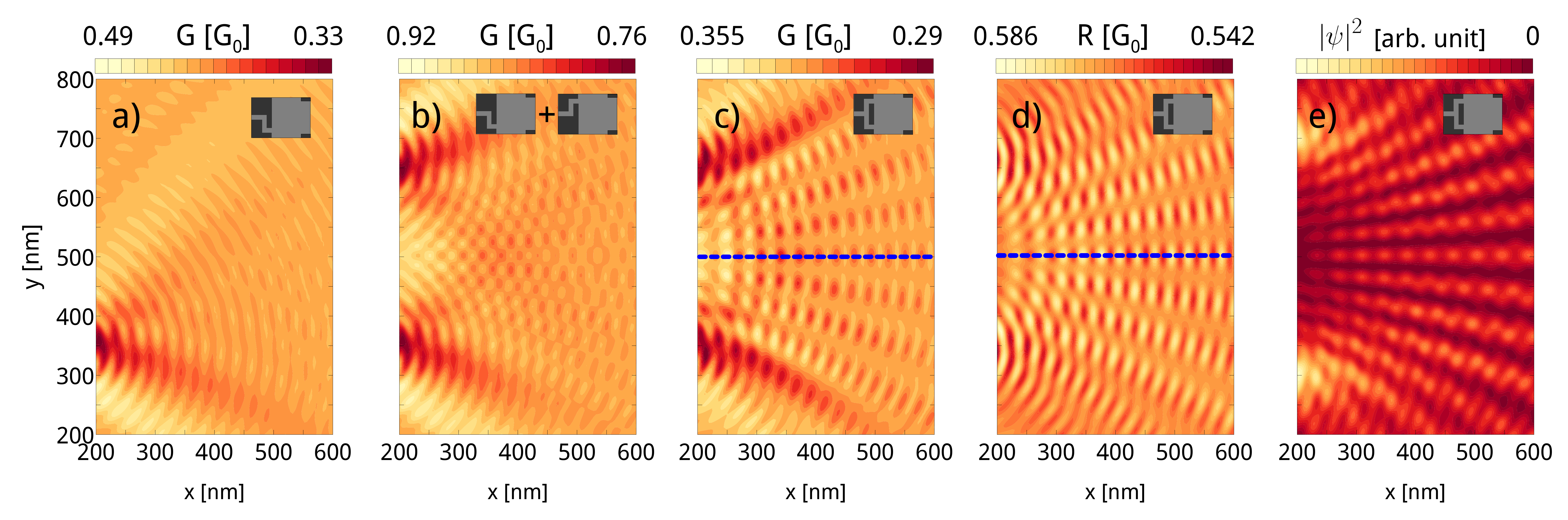} \caption{SGM images for the case of big value of $w_{\mathrm{qpc2}}=800$nm.
a) SGM-G image obtained for bottom slit open. b) Sum of images (a)
and image obtained for upper slit open (not presented here) c) The
SGM-G d) and SGM-R image obtained for both slits open. e) electron
density $|\psi|^{2}$ obtained for the system. Note the similarity
of (c), (d) and (e) images. The cross section of (c) and (d) along the blue line
is plotted in Fig. \ref{nowy}.}
\label{figszerokieQPC2}
\par\end{centering}
\end{figure*}

\begin{figure*}[htbp]
\begin{centering}
\includegraphics[width=0.4\paperwidth]{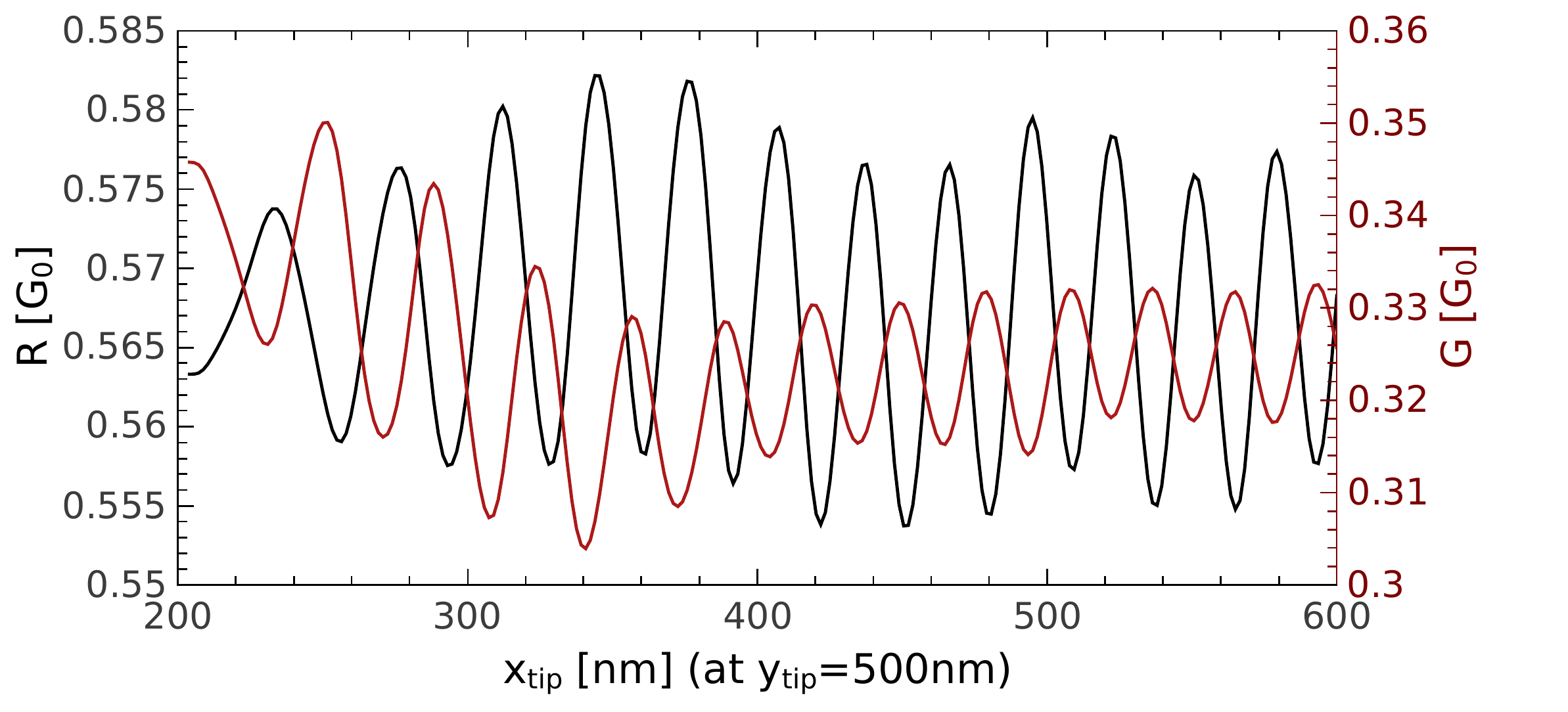}
\caption{Cross section of Fig. \ref{figszerokieQPC2} (c) and (d) taken along the blue line.
The black curve shows the backscattering $R$ (left axis), and the red one -- the conductance $G$ (right axis).}
\label{nowy}
\par\end{centering}
\end{figure*}

There is a way to restore the interference in the SGM-G images. Each
of the $M_{\mathrm{in}}$ subbands is with a certain probability backscattered,
transferred to QPC2 or goes out of the computational box through the
transparent boundary conditions to the rest of the system [see Eq.
(\ref{eqM})] and $M_{in}$ is independent of the tip position. Now
if we consider that $w_{\mathrm{qpc2}}$ becomes wider, the ratio
$G_{\mathrm{rest}}/G$ decreases, since QPC2 will be able to transfer
more current and the probability of the electron transfer form QPC1 to QPC2 increases
with the width of the latter. Thus for a large width $w_{\mathrm{qpc2}}$
the value of $G_{\mathrm{rest}}$ in Eq. (\ref{eqM}) will be small
and hence $\frac{2e^{2}}{h}M_{in}\approx R+G$. We can express the
conductance of the system in as $G\approx\frac{2e^{2}}{h}M_{in}-R\propto-R$.
We can see that for large values of $w_{\mathrm{qpc2}}$ the SGM-G
should start to exhibit the interference pattern as the SGM-R images
do. In Fig. \ref{figszerokieQPC2}(a-e) we show the results obtained
for a large value of QPC2 width $w_{\mathrm{qpc2}}=800$nm. Fig. \ref{figszerokieQPC2}(a)
shows the SGM-G image for the bottom slit open, and Fig.\ref{figszerokieQPC2}(b)
is a sum of the images obtained for bottom and upper slit open separately.
Now, the SGM-G image for both slits open (Fig. \ref{figszerokieQPC2}(c))
is clearly different from Fig. \ref{figszerokieQPC2}(b) with distinct
stripes due to the Young interference. We can compare this result
with SGM-R image of Fig.\ref{figszerokieQPC2}(d) which is now quite
similar to the SGM-G image, particularly in the center and close to
QPC2. In this region the approximated relation $G\propto-R$ is well
visible and SGM-G image is indeed a negative of SGM-R image.
To illustrate this further in Fig. \ref{nowy} we plotted cross section of these images
along the axis of the device.

Note that the increased width of QPC2 allows us to restore the interference
in the SGM-G images, but at the expense of the lost information on
the electron trajectories from QPC2 to QPC1. The double-slit interference
is present as long as one does not interfere with the measurement
trying to determine through which slit the particle passes \cite{Fey}.
Here, if we set the detector (QCP2) for mapping the electron trajectories
-- with a small value of $w_{\mathrm{qpc2}}$ -- we gain the information
about electron trajectories, but we lose the interference pattern.
Increasing the width $w_{\mathrm{qpc2}}$ leads to reduction of spatial
resolution of the detector, so we lose the paths in the images but
restore the interference pattern.

\subsection{Stability of classical trajectories in SGM-G images}

\begin{figure*}[htbp]
\begin{centering}
\includegraphics[width=0.4\paperwidth]{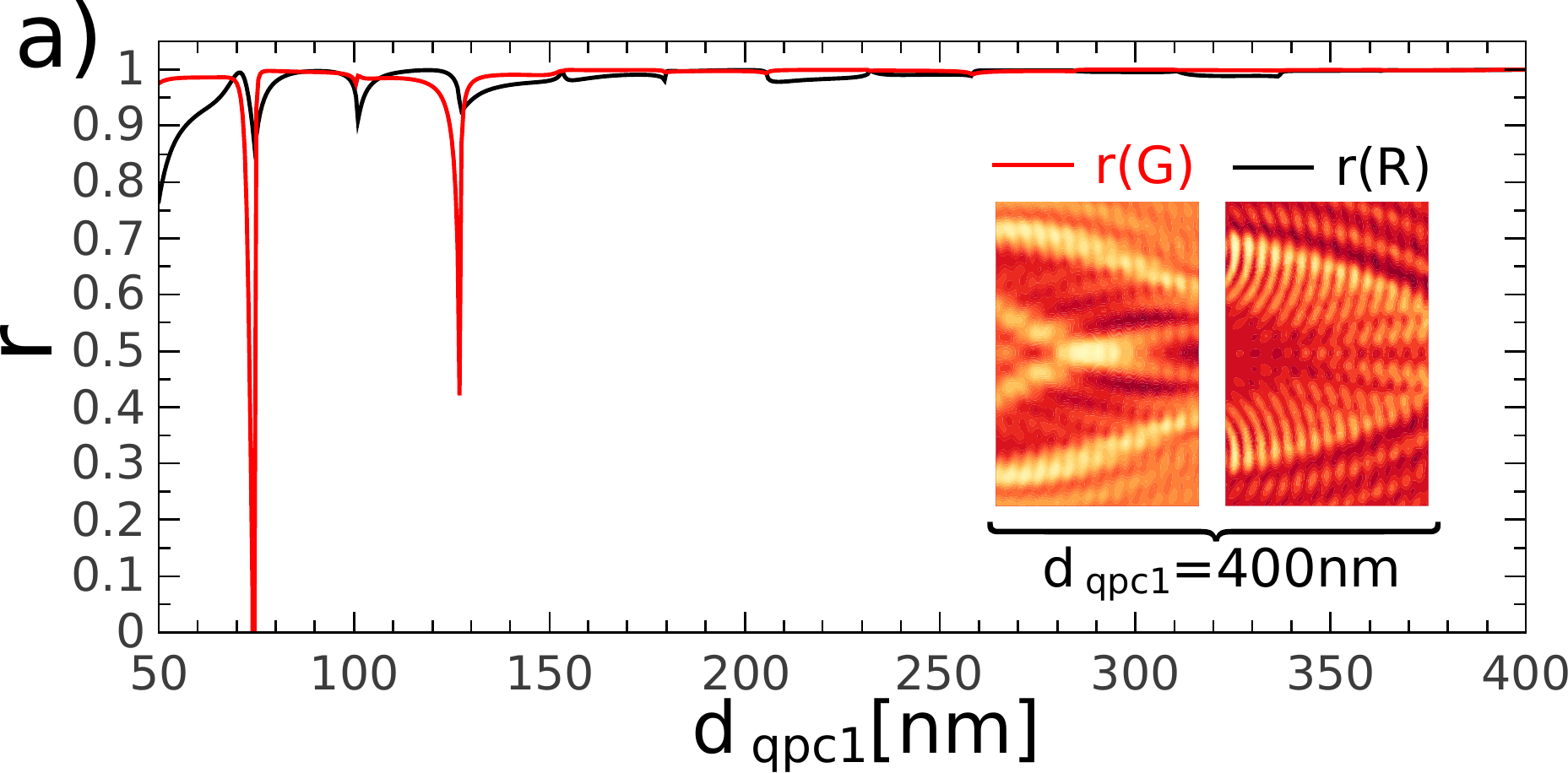}\includegraphics[width=0.4\paperwidth]{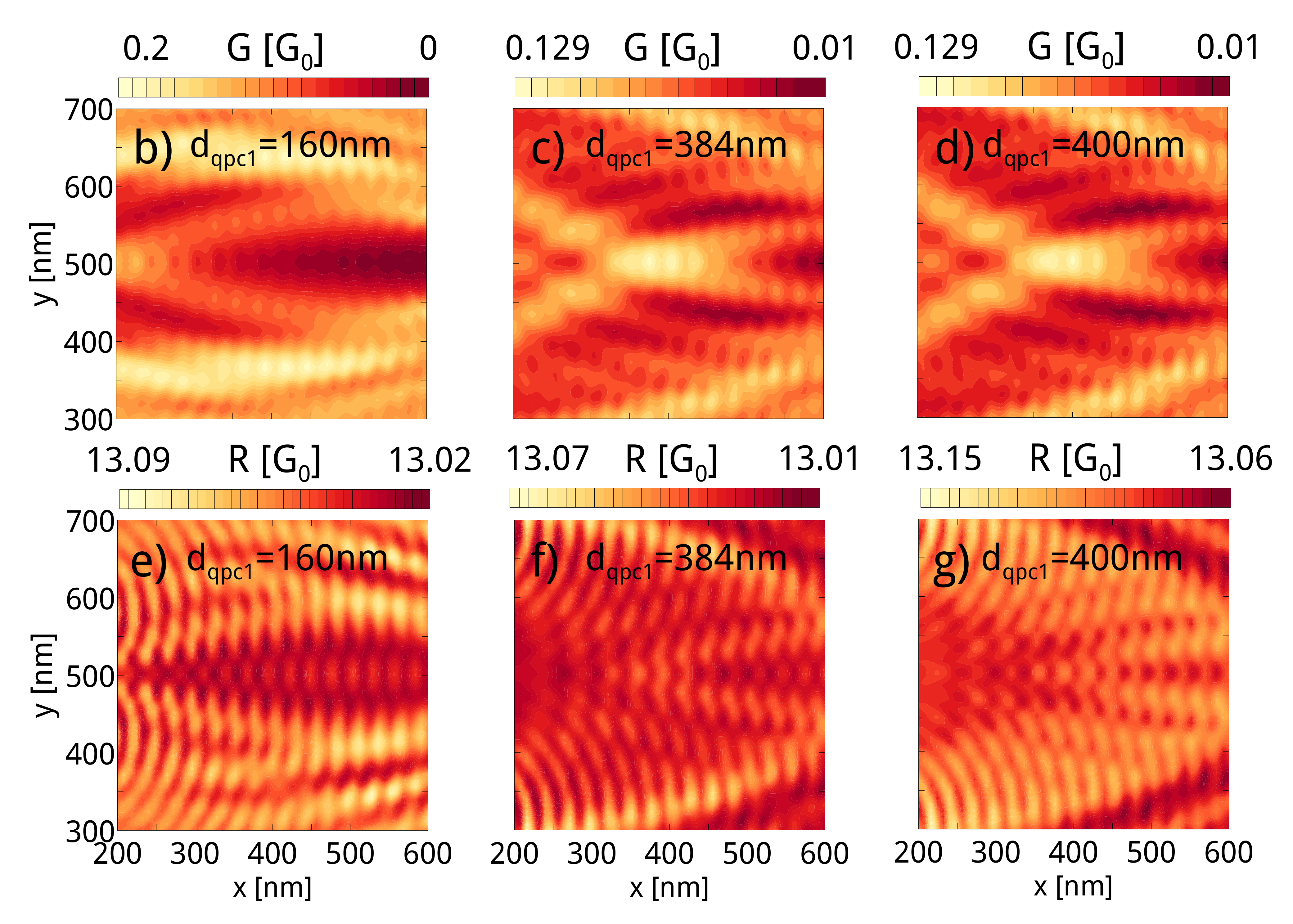}
\par\end{centering}

\caption{a) The Pearson correlation coefficient calculated between SGM-G/R
(r(G) and r(R)) images obtained for different values of $d_{\mathrm{qpc1}}$ the interslit distance [see Fig. 1, the width of the QPC1 slits is kept at 40 nm].
In the inset we show the SGM-G and SGM-R
reference images obtained for $d_{\mathrm{qpc1}}=400$nm. b-d) SGM-G
images obtained for different distances between the slits, see the
label inside each figure. e-g) Same as (b-d) but SGM-R images. The results were obtained for $U_{\mathrm{tip}}=15$ meV and $d_\mathrm{tip}=10$nm.}

\label{figstabilnosc}
\end{figure*}

The classical trajectories as extracted from the SGM-G images should
be stable against the geometrical parameters of the system, the distance between the QPC1 slits $d_\mathrm{qpc1}$ in particular.
For each value of the interslit distance $d_{\mathrm{qpc1}}$
we calculated the SGM-G($d_{\mathrm{qpc1}}$) and SGM-R($d_{qpc1}$)
images. The calculations were performed for $w_{\mathrm{qpc1}} = w_{\mathrm{qpc2}}=40$nm.
In Fig. \ref{figstabilnosc}(a) we plotted the Pearson correlation
coefficient \cite{kolasinski2013} between SGM-G($d_{\mathrm{qpc1}}$)
and last image SGM-G(400nm) -- the $r(G)$ curve, and correlation
between SGM-R($d_{\mathrm{qpc1}}$) and SGM-R(400nm) -- the $r(R)$
curve. We chose the SGM image for $d_{\mathrm{qpc1}}=400$nm as a
reference one. One may see in Fig. \ref{figstabilnosc}(a) that both
lines $r(G)$ and $r(R)$ quickly stabilize around 1 after $d_{\mathrm{qpc1}}=150$nm,
which means that through all the values from around $d_{\mathrm{qpc1}}=150$nm
to 400nm both images stay almost unchanged. Note, that the double-slit
interference in SGM-R images is only well resolved for the lowest
subband transport in the incident channel \cite{Kolasinski2014}.


We found that the SGM-G images are generally stable in function of
$d_{\mathrm{qpc1}}$, which means that paths are always visible, but
there is one exception when the distance between slits is small enough
such that both trajectories (from each QPC) overlap which makes the
calculated SGM-G maps difficult to interpret. Sample images of SGM-G
and SGM-R for small value of $d_{\mathrm{qpc1}}=160$nm are displayed
in Fig. \ref{figstabilnosc}(b) and (e), respectively. For larger
values of $d_{\mathrm{qpc1}}$ the classical trajectories are restored
(see Fig. \ref{figstabilnosc}(c) and (d)). Note that for Fig. \ref{figstabilnosc}(c)
and (d) the difference between the values of $d_{\mathrm{qpc1}}$
in each case is equal to 16 nm. The results for SGM-G are nearly identical.
However, the SGM-R images -- sensitive to the interference effects
-- very strongly depend on a specific value of $d_{\mathrm{qpc1}}$
-- cf. Fig. \ref{figstabilnosc}(f) and (g). The wave function passing
through both the input slits interferes with the AFM tip, as well
as with QPC2. The result of the interference in terms of the backscattering
depends on the variation of the distance between the slits of the
order of the period of the waves formed by interference at the Fermi level
which is equal to $\lambda_{\mathrm{F}}/2$.

\subsection{Wide tip potential}

\begin{figure*}[htbp]
\begin{centering}
\includegraphics[width=0.4\paperwidth]{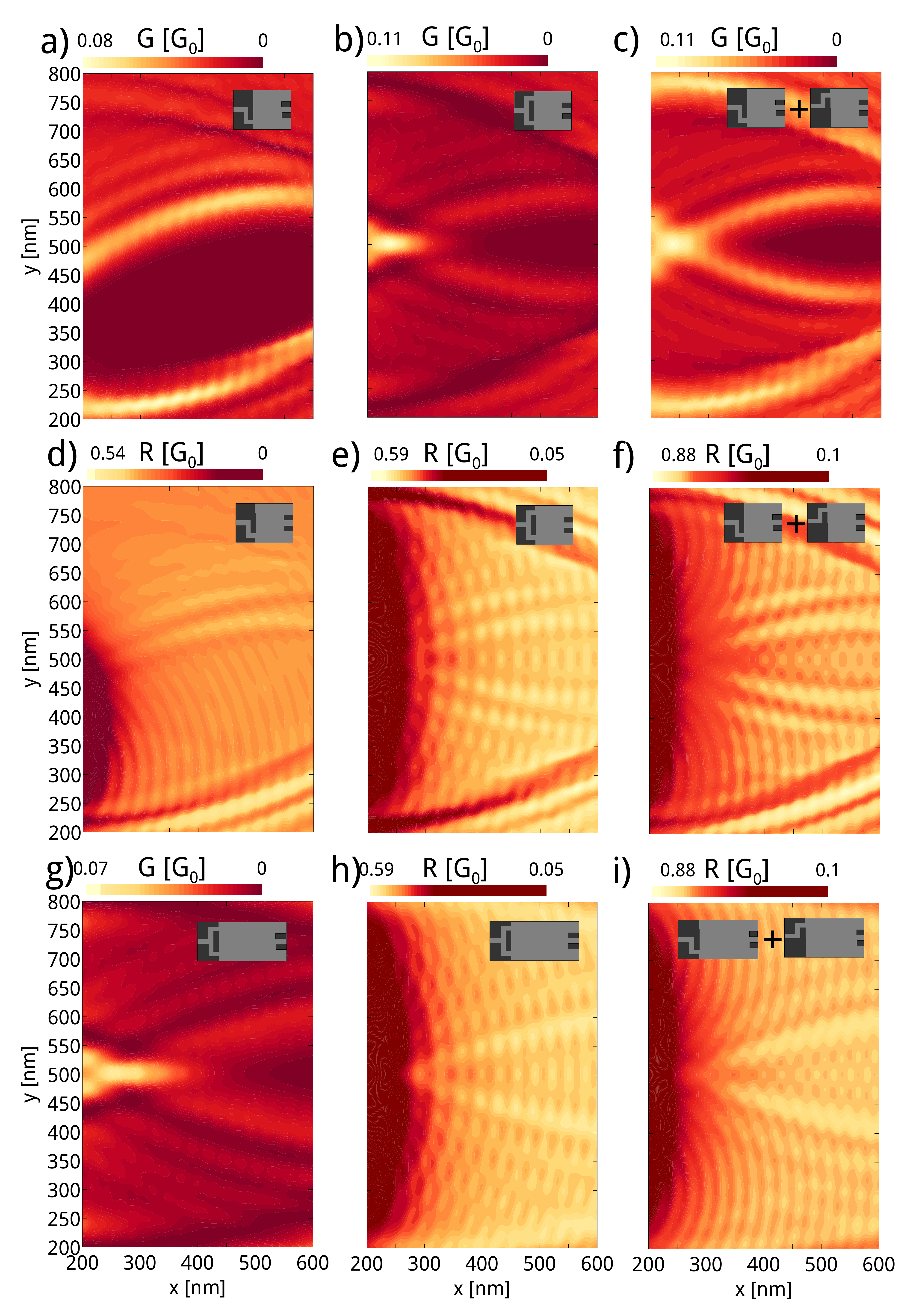}
\par\end{centering}

\caption{The SGM images obtained with $U_{\mathrm{tip}}=8$meV and $d_{\mathrm{tip}}=50$nm.
(a,d) SGM-G and SGM-R images obtained for lower slit open (b,e) the
same but for both slit open and (c,f) images obtained as a sum. (g-i)
Are the SGM images obtained in situation when the distance between
QPC1 and QPC2 was increased by 400 nm to 1 $\mu$m.
Panel (h) corresponds to backscattering in for both QPC1 slits open,
while in (i) a sum of signals for a single QPC1 slit open.}
\label{figdtip}
\end{figure*}

Let us return to the single subband transport in the input and the output
leads and consider the wider tip potential $d_{tip}=50$ nm. The results
for conductance $G$ for a single slit -- Fig. \ref{figdtip}(a) and
both the slits Fig. \ref{figdtip}(b) still indicate the classical
current paths, although naturally the width of the $G$ minima is
significantly increased. The $G$ map pattern is still very similar
to the one obtained by a sum of maps for separate slits [Fig. \ref{figdtip}]
- indicating a lack of double slit interference features in the source-drain conductance maps for $M_{out}=1$,
as discussed above.

The ''resistance'' map [Fig. \ref{figdtip}(e)] for both QPC1 slits open
contains i) the circular fringes near the input slits [single slit
interference of Fig. \ref{inter}(a)], ii) elliptical fringes [double
slit interference of Fig. \ref{inter}(d)], iii) the lateral fringes
[Fig. \ref{inter}(b)]. All the effects (i-iii) are tip-related
and in this case dominates  the Young interference --
which is weaker although still detectable. The Young interference
is restored when the QPC2 detector is placed further from the QPC1
[see Fig. \ref{figdtip}(h)], but naturally at the expense of
the amplitude of the $G$ signal [Fig. \ref{figdtip}(g)]. The
reduction of the lateral fringe pattern is observed for $R$ -- since
it involves backscattering by QPC2 -- is also observed [Fig. \ref{figdtip}(h-i)].

\section{Conclusions}

We have considered imaging of electron trajectories for
the double-slit experiment with the SGM technique.
Several interference mechanisms induced by the tip and the drain contact
as the electron detector have been found and the paths leading to the
interference have been identified. We studied the
SGM source-drain conductance maps and demonstrated that the classical
electron paths are clearly resolved but only for a narrow drain contact,
for which the double-slit interference features are absent. The double-slit
interference pattern is present in the conductance maps but only for
a wider drain contact, when the electron paths are no longer resolved.

We have indicated that a way to observe both trajectories and interference
pattern is to look simultaneously at two different SGM maps -- for
the backscattering $R$ -- which contain double-slit interference
signal and for the conductance $G$ -- that reveals the paths. The
latter allows one to map all the equivalent classical trajectories,
but without indicating the specific path the electron took on its
way to the drain channel.

\textbf{Acknowledgments} This work was supported by National Science
Centre according to decision DEC-2012/05/B/ST3/03290, and by PL-Grid
Infrastructure. The first author is supported by the scholarship of
Krakow Smoluchowski Scientific Consortium from the funding for National
Leading Reserch Centre by Ministry of Science and Higher Education
(Poland).


\begin{thebibliography}{10}
\bibitem{sgmr1} H. Sellier, B. Hackens, M.G. Pala, F. Martins, S.
Baltazar, X. Wallart, L. Desplanque, V. Bayot, and S. Huant, Sem.
Sci. Tech. \textbf{26}, 064008 (2011).

\bibitem{sgmr2} D.K. Ferry, A.M. Burke, R. Akis, R. Brunner, T.E.
Day, R. Meisels, F. Kuchar, J.P. Bird, and B.R. Bennett, Sem. Sci.
Tech. \textbf{26}, 043001 (2011).

\bibitem{topinka1} M. A. Topinka, B. J. LeRoy, S. E. J. Shaw, E.
J. Heller, R. M. Westervelt, K. D. Maranowski, and A. C. Gossard,
Science \textbf{289}, 2323 (2000).

\bibitem{topinka2} M. A. Topinka, B. J. LeRoy, R. M. Westervelt,
S. E. J. Shaw, R. Fleischmann, E. J. Heller, K. D. Maranowski, and
A. C. Gossard, Nature \textbf{410}, 183 (2001).

\bibitem{int1} A. Cresti, J. Appl. Phys. 100, 053711 (2006).

\bibitem{int2} M.P. Jura, M.A. Topinka, L. Urban, A. Yazdani, H.
Shtrikman, L.N. Pfeiffer, K.W. West, and D. Goldhaber-Gordon D, Nat.
Phys.\textbf{3}, 841 (2007).

\bibitem{int3} M. P. Jura, M. A. Topinka, M. Grobis, L. N. Pfeiffer,
K. W. West, and D. Goldhaber-Gordon, Phys. Rev. B \textbf{80}, 041303(R)
(2009).

\bibitem{int4} A. Abbout, G. Lemarie, and J.-L. Pichards, Phys. Rev.
Lett. \textbf{106}, 156810 (2011).

\bibitem{Jalabert2010}R. A. Jalabert, W. Szewc, S. Tomsovic, and
D. Weinmann, Phys. Rev. Lett. \textbf{105}, 166802 (2010).

\bibitem{Gorini2013}C. Gorini, R. A. Jalabert, W. Szewc, S. Tomsovic,
and D. Weinmann, Phys. Rev. B \textbf{88}, 035406 (2013).

\bibitem{Brun2014} B. Brun, F. Martins, S. Faniel, B. Hackens, G.
Bachelier, A. Cavanna, C. Ulysse, A. Ouerghi, U. Gennser, D. Mailly,
S. Huant, V. Bayot, M. Sanquer, and H. Sellier, Nat. Commun. 5, 4290
(2014)

\bibitem{kozikov2013}A. Kozikov, D. Weinmann, C. Rössler, T. Ihn,
K. Ensslin, C. Reichl, and W. Wegscheider, New J. Phys. \textbf{15},
013056 (2013).

\bibitem{paradiso} N. Paradiso, S. Heun, S. Roddaro, L.N. Pfeiffer, K.W. West, L. Sorba, G. Biasiol, F. Beltram, Physica E {\bf 42}, 1038 (2010)

\bibitem{Kolasinski2014}K. Kolasinski, B. Szafran, M. P. Nowak, Phys.
Rev. B. \textbf{90}, 165303 (2014)

\bibitem{Fey} R.P. Feynman, Robert B. Leighton, and Matthew Sands
\textit{The Feynman Lectures on Physics,} Vol. 3., Addison-Wesley
(1965); R. Bach, D. Pope, S.H. Lou, and H. Batelaan, New J. Phys.
\textbf{15}, 033018 (2013).

\bibitem{crook} R. Crook, C.G. Smith, M.Y. Simmons, and D.A. Ritchie,
Phys. Rev. B \textbf{62}, 5174 (2000).

\bibitem{skipping} K. E. Aidala, R. E. Parrott, T. Kramer, E. J.
Heller, R. M. Westervelt, M. P. Hanson, and A. C. Gossard, Nat. Phys.
\textbf{3}, 464 (2007).

\bibitem{aidala} K. E. Aidala, R. E. Parrott, E.J. Heller, and R.M.
Westervelt Physica E, \textbf{34}, 409 (2006).

\bibitem{khatua2014} P. Khatua, B. Bansal, and D. Shahar, Phys. Rev.
Lett. 112, 010403 (2014)

\bibitem{kozikov} A. Kozikov, D. Weinmann, C. Rössler, T. Ihn, K.
Ensslin, C. Reichl, and W. Wegscheider, New J. Phys. \textbf{15},
083005 (2013).

\bibitem{kozikov2}A. Kozikov, R. Steinacher, C. Rössler, T. Ihn,
K. Ensslin, C. Reichl, and W. Wegscheider, New. J. Phys. \textbf{16},
053031 (2014).

\bibitem{kolasinski2013}K. Kolasi\'{n}ski, B. Szafran Phys. Rev.
B \textbf{88}, 165306 (2013)


\bibitem{szafran} B. Szafran, Phys. Rev. B \textbf{84}, 075336 (2011)

\bibitem{ensliny} N. Pascher, F. Timpu, C. R\"ossler, T. Ihn, K. Ensslin, C. Reichl, and W. Wegscheider,
Phys. Rev. B {\bf 89}, 24508 (2014).

\bibitem{martinsy} F. Martins, S. Faniel, B. Rosenow, M. G. Pala, H. Sellier, S. Huant, L. Desplanque, X. Wallart, V. Bayot, and B. Hackens, New J. Phys. {\bf 15}, 013049 (2013).

\bibitem{atleast} B. Hackens, F. Martins, S. Faniel, C. A. Dutu,
H. Sellier, S. Huant, M. Pala, L. Desplanque, X. Wallart, V. Bayo,
Nature Commun. \textbf{1}, 39, (2010)

\bibitem{Nowak2014} M. P. Nowak, K. Kolasinski, B. Szafran, Phys.
Rev. B \textbf{90}, 035301 (2014)

\bibitem{Kolasinski2014QHI}K. Kolasinski, B. Szafran, Phys. Rev.
B \textbf{89}, 165306 (2014)

\bibitem{Lent90}D. J. Kirkner and C. S. Lent, Journal of Applied
Physics \textbf{67}, 6353 (1990)

\bibitem{Lent94}M. Leng and C. S. Lent, Journal of Applied Physics
76, 2240 (1994)

\bibitem{Bagwell1990}P. F. Bagwell, Phys. Rev. B \textbf{41}, 15
(1990)

\bibitem{Zwierzycki2008}M. Zwierzycki, P. A. Khomyakov, A. A. Stariko
K. Xia, M. Talanana, P. X. Xu, V. M. Karpan, I. Marushchenko, I. Turek,
G. E. W. Bauer, G. Brocks, and P. J. Kelly, Physica Status Solidi
B \textbf{245}, issue 4 p. 623-640 (2008) \end{thebibliography}
\end{document}